\documentclass[12pt]{article}
\usepackage{psfrag}
\usepackage[top=23mm,bottom=20mm,left=25mm,right=20mm]{geometry}
\usepackage{amsmath,amsthm}
\usepackage{graphicx}
\usepackage{amssymb}
\usepackage{amsopn}
\usepackage{cite}

\newcommand{\ld}{\lambda}
\newcommand{\lda}{\lambda^{-\frac{1}{2}}}
\newcommand{\ldb}{\lambda^{\frac{1}{2}}}
\newcommand{\daa}{\frac{\dot a}{a}}
\newcommand{\dll}{\frac{\dot \lambda}{\lambda}}
\newcommand{\dff}{\frac{f''(R)\dot R}{f'(R)}}
\newcommand{\dpp}{\frac{\dot \phi}{\phi} }
\newcommand{\ddaa}{\frac{\ddot a}{a}}
\newcommand{\ddll}{\frac{\ddot \lambda}{\lambda}}
\newcommand{\ddff}{\frac{f'''(R)\dot R^2+f''(R)\ddot R}{f'(R)}}
\newcommand{\ddpp}{\frac{\ddot \phi}{\phi}}
\newcommand{\lp}{\lda\phi^{-1}}
\newcommand{\pot}{\left(\frac{\phi}{\alpha m}\right)
          ^{\frac{1}{m-1}}}

\begin{document}

\title{\bf Five-dimensional metric $f(R)$ gravity\\
       and the accelerated universe}
       \author{Biao Huang\footnote{E-mail:
       phys.huang.biao@gmail.com},
        Song Li and Yongge Ma\footnote{Corrsponding author; E-mail: mayg@bnu.edu.cn}\\
       Department of Physics, Beijing Normal University,
       Beijing 100875, China}
\date{}
\maketitle

\begin{abstract}
 The metric
  $f(R)$ theories of gravity are generalized to
 five-dimensional spacetimes. By assuming a hypersurface-orthogonal
  Killing vector field
 representing the compact fifth dimension, the five-dimensional
 theories are reduced to their four-dimensional formalism.
 Then we study the cosmology of a special class of
 $f(R)=\alpha R^m$ models
  in a spatially flat FRW spacetime. It is shown that
 the parameter $m$ can be constrained to a certain range by the
 current observed deceleration parameter, and its lower bound corresponds
 to the Kaluza-Klein theory. It turns out that both
 expansion and contraction
 of the extra dimension may prescribe the smooth transition from the
 deceleration era to the acceleration era in the recent past as well as an accelerated scenario for
 the present universe. Hence
 five-dimensional $f(R)$ gravity can naturally account for the
 present accelerated expansion of the universe.
   Moreover, the models predict a transition
 from acceleration to deceleration in the future, followed by
 a cosmic recollapse within finite time.
  This differs from the
 prediction of the five-dimensional Brans-Dicke theory
 but is in consistent with a recent prediction based on loop quantum
 cosmology.


\end{abstract}

\section{Introduction}

  Since 1998, a series of independent observations,
 including type Ia supernova
  \cite{ia1,ia2,ia3}, weak lens \cite{wl},
 cosmic microwave background
 anisotropy \cite{cmb},
 large scale structure \cite{lss1,lss2}, baryon oscillation
 \cite{bo1,bo2},
   etc., have confirmed that
 our universe is undergoing a period of accelerated expansion.
 This result conflicts fiercely with the prediction of
 general relativity, and therefore triggers a ``golden age'' of
 cosmology since researchers in cosmology, astrophysics, general
 relativity, particle physics are all involved in seeking for
 viable models to explain this phenomena.
 Within the frame of general relativity, the cosmic speed-up can be
 viewed as an indication that the present
 universe is dominated by certain mysterious fluid with large
 negative pressure, called "dark energy". However, such simple
 explanations could hardly be satisfactory. The simplest
  $\Lambda$CDM model, which fits the observation data best so far,
  assumes a cosmological constant to be
 responsible for the dark content. But the observed value of the
 constant $\Lambda$ is unnaturally smaller than any estimation by tens
 of orders \cite{toolarge1, toolarge2}. Other dynamical dark
 energy models then waive evolving scalar fields to
 replace and resemble the cosmological constant (for review see
 \cite{de}). These early attempts provide us with initial
 understanding of the complexity of the problem, though they
  chiefly serve for empirical fittings with comparatively
  poor theoretical
   motivation, and none of such models turn out to be
 problem-free.

 In view of the challenge and the fact that gravity is the only
 dominant long-range
 interaction insofar as we know, it is reasonable to consider
 the possibility that we have not fully understood the character of
 gravity on a cosmological scale. Therefore, the exploration of
 alternative
 theories
 of gravity are proposed.
 One of such examples
  is the Kaluza-Klein (KK) theory, which was initiated
 by the motivation of unifying the
 gravitation field and the electromagnetic field in a
 five-dimensional (5-D) metric.
 Since the fifth dimension is supposed to be a compact $S^1$
 circle with an extremely tiny radius, it would actually
 yield no observable effect.
 Graceful though, the original version of KK theory
  fails in passing the
 Solar-System experiment \cite{xupengppn}. But recently a new model
 of modified (non-compact) Kaluza-Klein cosmology was investigated
 in \cite{noncompactkk}, where the universe in turns
  inflates,
 decelerates, and then accelerates in respectively early times,
 radiation dominated era and matter dominated era. This result
 meets the observational facts roughly. But there still
 exist problems such as that it leaves no decelerated period in the matter-dominant
 era for the formation of large scale structure.
  Besides, in \cite{4dbd} Brans-Dicke (BD) theory of gravity was
  introduced to match the Mach's principle. It is
   found in \cite{5dbd} that 5-D BD theory
  can naturally predict the cosmic acceleration
  without
  the requirement of a time-varying BD parameter $\omega$ or a
  fabricated
  potential in
   its 4-D counterpart
  \cite{bddisadv1,bddisadv2}.
   Thus, it lends
  confidence and motivation
   to the effort of extending gravitational theories in
  4-D to 5-D space-time.
   Also, note that in \cite{5dbd} the allowed range
  for parameters $\omega-n$ is considerably wide, where $\omega$ is
  the BD parameter and $n$ represents the interaction between
  the extra dimension and other spatial ones.
  We may suspect that further information
  concerning the different choices of $\omega-n$ are masked in the
  formalism of 5-D BD theory. Given the argument for the
  equivalence (by evolutional effect)
  of the $f(R)$ theory of gravity to
  certain special cases of BD theory up to
  a potential term \cite{bdfr}, we are led to
  consider the 5-D $f(R)$ theories of gravity for such information.
  It is worth noting that the scale of the compact extra dimension of 5-D
  gravity theories, which we are considering, is constrained to be
  less than 50 $\mu m$ by current experiments \cite{kapner,tu,turyshev}.

 The $f(R)$ theories of gravity extends the
 Hilbert-Einstein action in general
 relativity to
 \begin{equation}
   S = \int d^4 x \sqrt{-g} f(R),
 \end{equation}
 where $f(R)$ is an arbitrary function of the curvature scalar $R$. The
 reason for such exploration in 4-D spacetime
  is twofold. On the one hand, recent
 research has shown that a small correction to the Hilbert-Einstein
 action by adding an inverse term of $R$
 would lead to acceleration of the
 universe \cite{cdtt}, and a large variety of models are proposed
 since then (for review paper, see \cite{frreview}). However,
 few of them turn out to be without problem and most of them
 admit  fine-tunings so as to be consistent with experiments.
  On the other hand, it is highly
 possible that future quantum theory of gravity, such
 as string theory, loop quantum gravity, etc., would bring about
 modifications to the action for
 classical general relativity, just like that
 general relativity has introduced corrections to the
 Newton's gravitation theory. Therefore, this effort would potentially
  match the low energy effective theory of
   quantum gravity and therefore accelerate its birth.

 With above motivations,
  we thought that modified gravity without fine-tuning
  should better be constructed in higher dimensions.
  It is shown in this paper
 that values of the parameters
 in the 5-D theory are all with
 reasonable deductions and interpretations.
 In 5-D spacetimes with a hypersurface-orthogonal Killing vector field, the 5-D $f(R)$
  theories of gravity are
 reduced to the sensible 4-D world in section
 \ref{killing}.
  Here we have
 adopted KK's idea that the fifth dimension is a small
  unobservable
  compact ring $S^1$, so
 that a Killing vector field would arise naturally in the low energy
 environment.
 Further, in view of that in KK's theory one
 can find the sensible 4-D universe to be expanding (without
 acceleration) as a result of the contraction of the extra
 dimension, we assume that
 there is certain interaction between the
 fifth dimension and other ones at present, and take their
 relationship to be the general power-law form.
 Then the reduced theories are carefully studied in the $k=0$ case of the FRW metric
 in section \ref{frw}.
 We consider a special class of models with
 $f(R)=\alpha R^m$ and
  present the numerical simulation of the evolution history
 of the matter dominated universe in section \ref{numeri},
 demonstrating its consistency with the observation.
 Finally, we end with discussions in section \ref{jielun}
  on the similarity and difference between
 the 5-D $f(R)$ theories
 and the 5-D BD theory, as well as
 further indications from our 5-D models.

 \section{Killing reduction of the 5-D $f(R)$ theory}\label{killing}

 We start with the action in 5-D space-time:
 \begin{equation}\label{action}
   S = \frac{1}{2\kappa} \int d^5 x \sqrt{-g} f(R) \, + S_M.
 \end{equation}
 Here $\kappa= 8\pi G^{(5)}  /c^4$,
  where $G^{(5)}$ denotes the gravitational
 constant in the 5-D spacetime, and $S_M$ represents the matter term
 in the total action $S$. Variation of (\ref{action})
  with respect to the metric $g^{ab}$ gives
 \begin{equation}\label{5dfield}
   f'(R)R_{ab} - \frac{1}{2} g_{ab}f(R) - (\nabla_a\nabla_b-
   g_{ab}\nabla^c\nabla_c)f'(R) = \kappa T_{ab},
 \end{equation}
 where $T_{ab}=\frac{-2}{\sqrt{-g}} \frac{\delta S_M}{\delta
 g^{ab}}$, and $f'(R)=\mbox{d} f(R)/\mbox{d} R$.
 Note that we have employed the abstract index
 \cite{wald} and $R, R_{ab}, T_{ab}$
 represent quantities in the 5-D spacetime.
  Contracting equation (\ref{5dfield}) with $g^{ab}$, we obtain
 the dynamical equation for the scalar field $f'(R)$:
 \begin{equation}
   \label{5ddynamics}
   \nabla^a\nabla_a f'(R) = \frac{1}{4} [ \kappa T - R f'(R) +
   \frac{5}{2} f(R)].
 \end{equation}
 Since here $g_{ab}g^{ab}$=5 instead of 4, the dynamical
 equation of $f'(R)$ differs from its usual 4-D counterpart.

 Next we consider the structure of the 5-D spacetime and its
 relation to our sensible 4-D world.
 First, as mentioned in the introduction, we assume that
  the 5-D spacetime possesses a Killing vector field $\xi^a$ which
  represents the fifth dimension and is everywhere space-like.
  Recall that the Killing
  reduction of 4-D spacetime is studied by Geroch \cite{4dreduction}
  and is further extended to the 5-D spacetime by Yang {\em et al.}
  \cite{5dreduction}.
  Therefore, following these works
  we introduce the 5-D metric as
  \begin{equation}
    g_{ab}=h_{ab}+\lambda^{-1}\xi_a\xi_b,
  \end{equation}
  where $h_{ab}$ is the metric in the usual 4-D universe and $\lambda=
  \xi^a\xi_a$. If $\xi^a$ is not hypersurface-orthogonal, it would be related to
   electromagnetic 4-potential in the reduced 4-D theory,
   as demonstrated by KK theory.
  Since we are chiefly concerned with the cosmological effect of
  the reduced model, our discussion is restricted to the case where
  $\xi^a$ is hypersurface-orthogonal for convenience.
   Without losing generality, a coordinate system
  can be chosen as $\{x^\mu, x^5\}, \mu=0,1,2,3$, with
  $(\frac{\partial}{\partial x^5})^a = \xi^a$. Then the line element
  of $g_{ab}$ reads $ds^2 = g_{\mu \nu} dx^\mu dx^\nu + \lambda
  dx^5 dx^5$, while Ricci tensors
   and the dynamical equation of $\lambda$ in 5-D and 4-D spacetime
   bear the following relation \cite{5dreduction}:
 \begin{equation}\label{riccireduction}
   R_{ab}^{(4)} = \frac{1}{2} \lambda ^{-1} D_a D_b \lambda -
   \frac{1}{4}\lambda^{-2} (D_a\lambda)D_b \lambda + h_a^m h_b^n
   R_{mn},
 \end{equation}
 \begin{equation}
    \label{lambdareduction}
   D^aD_a\lambda = \frac{1}{2}\lambda^{-1}(D^a\lambda) D_a \lambda -
   2R_{ab}\xi^a\xi^b.
 \end{equation}
 Here $D_a$ denotes the covariant derivative on 4-D spacetime
  and is defined as \cite{wald}
  \[D_a T_{b_1 \dots b_n}^{c_1 \dots c_m} = h_a^d
 h_{b_1}^{e_1}\dots h_{b_n}^{e_n} h^{c_1}_{f_1} \dots
 h^{c_m}_{f_m} \nabla_d T_{e_1 \dots e_n}^{f_1\dots f_m},\]
 satisfying all the conditions for a derivative operator.
 On the other hand, the stress-energy tensor in
 (\ref{5dfield}) is regarded as a perfect fluid in 5-D spacetime with
 the expression \cite{5dbd}
 \begin{equation}\label{tab}
    T_{ab}^{(5)} = L^{-1}\lda \left[ (\rho + P) U_a U_b + P
    g_{ab}\right] =
    L^{-1}\lda\left[T_{ab}^{(4)} + P \ld \xi_a\xi_b\right],
  \end{equation}
 where $\rho$ and $P$ are the 4-D
 energy density and the hydrostatic pressure respectively, and L is
 a constant
 representing the coordinate scale of the fifth dimension.
 This extension of the stress-energy tensor of the 4-D perfect fluid
  to a 5-D one can be understood in the following way:
    because
  the fifth dimension is compact and attached to every point of the
 4-D space-time, $\rho$ and $P$ experienced by a 4-D observer
 should be an integrated effect throughout
  the compact ring. If we further expect that
   the fluid distributes homogeneously
  and does not travel along the fifth
  dimension,
 it is clear that
 $\rho=\int \rho^{(5)}\lambda^{1/2}dx^5=\rho^{(5)}\lambda^{1/2}
 L$, $P=P^{(5)}\lambda^{1/2}L$, and hence
  equation (\ref{tab}) is obtained.
 Combining (\ref{5dfield}), (\ref{5ddynamics}) with
 (\ref{riccireduction}),
 (\ref{lambdareduction}), and (\ref{tab})
 , straightforward
 calculations lead to the 4-D field equation:
 \begin{eqnarray}\nonumber
   \label{4dfield}
   G_{ab}^{(4)} &=& \frac{8\pi G}{c^4}\frac{\lda}{f'(R)} T_{ab}^{(4)}
     +   h_{ab}\frac{f(R)-R f'(R)}{2f'(R)}\\ \nonumber
     &&+ \frac{1}{2} \lambda^{-1} (D_a D_b -h_{ab}D^cD_c)\lambda
     - \frac{1}{4}\lambda^{-2} [ (D_a\ld)D_b\ld - h_{ab}(D^c\ld)
     D_c\ld]\\
     &&+ \frac{1}{f'(R)} (D_a D_b -h_{ab}D^c D_c)f'(R)
     - \frac{1}{2 f'(R)}\lambda^{-1} h_{ab} (D^c\ld)D_cf'(R),
 \end{eqnarray}
 and the dynamical equations of $\lambda$ and $f'(R)$:
 \begin{eqnarray}\nonumber
   \label{4dlambda}
   D^aD_a\ld & =& \frac{8\pi G}{c^4} \frac{\ldb}{f'(R)}
          \left(\frac{1}{2} T^{(4)} - \frac{3}{2} p\right)
     + \frac{1}{2} \lambda^{-1}(D^a\ld)D_a\ld -
     \frac{1}{f'(R)} (D^a\ld)D_af'(R)\\
     &&+ \frac{\ld}{f'(R)}\left[\frac{1}{4}f(R)
       -\frac{1}{2}R f'(R)\right],
 \end{eqnarray}
 \begin{equation}
   \label{4dfr}
   D^aD_a f'(R) = \frac{8\pi G}{c^4} \frac{\lda}{4} (T^{(4)} +p)
      -\frac{1}{2}\ld^{-1}(D^a\ld)D_af'(R)
      +\frac{1}{4}\left[\frac{5}{2}f(R) - R f'(R)\right],
 \end{equation}
 where $G=G^{(5)} L^{-1}$ represents the usual 4-D gravitational constant
 \cite{4d5dg}.

 \section{FRW cosmology of the reduced $f(R)$ gravity}\label{frw}

 In this section, we will study the cosmological predictions of
 the reduced $f(R)$ gravity. The metric of
  a spatially isotropic and homogeneous
 4-D spacetime is the
 Friedman-Robertson-Walker (FRW) metric with three possible
 structures  of the space.
 As suggested by the observation \cite{k=0},
  we only handle the spatially flat case
 where the 4-D line element reads
 \begin{equation}
   ds^2 = -dt^2 + a^2(t) \sum\limits_{i=1}^{3}dx_i^2.
 \end{equation}
 Then the two components of the field equation (\ref{4dfield}) are:
 \begin{equation}
   \label{gab1}
   3\left(\daa\right)^2 = 8\pi G \frac{\lda}{f'(R)} \rho
        -\frac{3}{2} \daa \dll - 3\daa \dff - \frac{1}{2} \dll\dff
        -\frac{f(R)-R f'(R)}{2f'(R)}
 \end{equation}
 and
 \begin{eqnarray}\label{gab2}\nonumber
   -\left( 2\ddaa + \frac{\dot a^2}{a^2}\right) &=&
        \frac{8\pi G}{c^2} \frac{\lda}{f'(R)} P
        +\daa\dll -\frac{1}{4}\left(\dll\right)^2
        +\frac{1}{2}\dll\dff +2\daa\dff\\
        &&+\ddll
        +\ddff
        +\frac{f(R)-R f'(R)}{2f'(R)}.
 \end{eqnarray}
 The dynamical equations of $\ld$ and $f'(R)$ are respectively:
 \begin{equation}
   \label{lambda1}
   \ddll = 8\pi G \frac{\lda}{f'(R)}\left(\frac{\rho}{2}\right)
       -3\daa\dll + \frac{1}{2}\left(\dll\right)^2
       -\dll\dff -\frac{f-2R f'(R)}{4f'(R)},
 \end{equation}
 and
 \begin{equation}
   \label{phi1}
   \ddff = 8\pi G \frac{\lda}{f'(R)}
      \left(\frac{\rho}{4} - \frac{P}{c^2}\right)
      -3\daa\dff - \frac{1}{2}\dll\dff
      -\frac{  \frac{5}{2}f(R) - R f'(R) }{4f'(R) }.
 \end{equation}
 Here an over-dot denotes the derivative with respect to the
 proper time $t$ of the isotropic observer, and we have assumed that
 the scalar field $\ld$ depends only on $t$.
  Note that the conservation equation of the 5-D stress-energy
  tensor, $\nabla^\mu T_{\mu 0} = 0$, gives:
  \begin{equation}
    \label{con}
    \dot \rho + 3 \daa (\rho + \frac{P}{c^2})
    + \frac{1}{2}\dll \frac{P}{c^2} = 0,
  \end{equation}
  which can also be obtained directly from equations
  (\ref{gab1}) -- (\ref{phi1}).
  For the common baryonic matter content of the
  present universe, the pressure is negligible compared to the
  energy density. Thus, we set $P/c^2=0$ by approximation
    in the above equation and get
   $\rho = \rho_0 (a_0 / a )^3$,
   where $\rho_0$ is the current observed energy density of the luminary
    matter.

 By now, we have been
  dealing with the general form of $f(R)$ gravity. Next
  we restrict our discussion to a specific
 class of $f(R)$ models, $f(R)=\alpha R^m$,
  to further study the evolutional
 characteristics of the theory. It has been shown that this choice in 4-D spacetime,
 with $m\neq1$ and the common baryonic matter, could fit the
 Hubble diagram of Type Ia supernovae without need of dark energy \cite{form}.
 We denote $f'(R) = \alpha m R^{m-1} \equiv
 \phi $ and obtain
 \begin{equation}
 \label{rphi}
  R = \left(\frac{\phi}{\alpha m}\right)
 ^{\frac{1}{m-1}}, \;\;\;\;\; f(R)= \frac{\phi}{m}
 \left(\frac{\phi}{\alpha m}\right)^{\frac{1}{m-1}},
 \end{equation}
 where $\alpha$ is a dimensional constant.
 With the above analysis, we find the three cosmological
  evolution equations for numerical simulations
   from the combination of equations
   (\ref{gab1}) -- (\ref{phi1}):
  \begin{equation}
    \label{ai}
    \ddaa =
        \left(\daa\right)^2
        +\daa\dll + 2\daa\dpp + \frac{1}{2}\dll\dpp
        + \left(\frac{3}{8m} - \frac{1}{4}\right)\pot
        - 8\pi G \lp
        \frac{3}{4}\rho_0\left(\frac{a_0}{a}\right)^3,
  \end{equation}
  \begin{equation}
    \label{lambdai}
    \ddll =
        -3\daa\dll + \frac{1}{2}\left(\dll\right)^2 - \dll\dpp
        - \left(\frac{1}{4m} - \frac{1}{2}\right)\pot
        + 8\pi G \lp \frac{1}{2}\rho_0\left(\frac{a_0}{a}\right)^3,
  \end{equation}
  \begin{equation}
    \label{phii}
    \ddpp =
        -3\daa\dpp - \frac{1}{2}\dll\dpp
        - \left(\frac{5}{8m} - \frac{1}{4}\right)\pot
        + 8\pi G \lp \frac{1}{4}\rho_0\left(\frac{a_0}{a}\right)^3.
  \end{equation}

 \section{Numerical simulations}
 \label{numeri}

\begin{figure}
 \begin{center}
 \includegraphics[height=7cm, width=13cm]{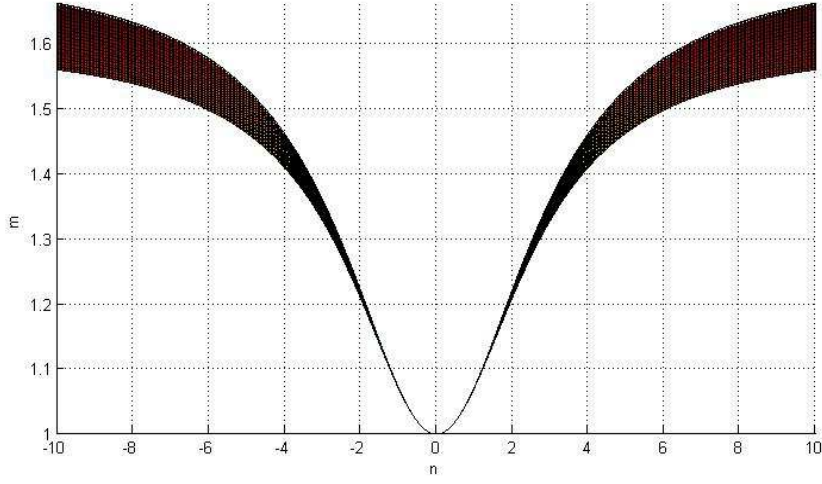}
 \caption{\label{range1} Allowed range for the values of m and
 n with $q_0 \in (-0.47, -0.67)$ \cite{experiment}.
 The horizontal axis stands for different value of $n$, while
 the vertical axis represents $m$.  }
 \end{center}
 \end{figure}

 Let us now find the natural initial values of $a_0, \dot{a_0},
 \ld_0, \dot{\ld_0}, \phi_0,  \dot\phi_0$
 for the numerical simulation.
 Firstly,
 $a_0, \ld_0$ themselves have no direct
 physical meaning and therefore can be
 simply fixed as 1 (with no dimension). Then we can directly
  determine the value
 of $\dot a_0$ by the present value of Hubble parameter
 $H_0=(\dot a/a)_{t_0}$.
 Secondly,
  we adopt the dynamical compactification idea of KK cosmology that the contraction of the
  extra dimension would result in the expansion of the remaining
  dimensions \cite{Appelguist}, and assume
  \begin{equation}
    \label{kk}
    a^3 \ld ^{n/2} = constant,
  \end{equation}
  where for a particular $f(R)$ theory with given $m$ and $\alpha$ in Eq.(\ref{rphi}),
  the free parameter $n$ can be constrained by the
  observation. Therefore we have
  $(\dot \ld / \ld )_{t_0}=6H_0 /n$.
    Finally,
  comparing the matter term in
 equation (\ref{4dfield}) to the field equation of general
 relativity, we expect $\ld ^{-1/2} \phi^{-1} \sim 1$, at least
 for the present period; thus $\phi_0=1$ (with no dimension
 either) and further
$ \left(\dpp\right)_{t_0}=-\frac{1}{2}\left(\dll\right)_{t_0}$.
 In summary, the initial conditions are:
 \begin{equation}\label{ini}
 \left\{\begin{array}{l}
 a_0 = \ld _0 = \phi_0 = 1,\\
 \dot a_0 = H_0, \, \dot \ld _0 = -\frac{6}{n}H_0, \,
 \dot \phi_0 = \frac{3}{n}H_0.
 \end{array}
 \right.
 \end{equation}
 Then  we
 estimate the value
  of $\alpha$ in equations (\ref{ai}) -- (\ref{phii}) and
  determine the allowed range for the pair of parameters $m$
  and $n$.
 A calculation of the 5-D curvature scalar shows
 \begin{equation}
   \label{5dr}
   R = 6\left(\ddaa+\frac{\dot a^2}{a^2}\right)
        + \ddll + \frac{1}{2}\left(\dll\right)^2 + 3\daa\dll.
 \end{equation}
 Inserting (\ref{rphi}),
 (\ref{lambdai}) into the expression of $R$ and
 noting that the deceleration parameter $q=-\frac{\ddot a a}{a}$
 is an observable with its present value denoted as $q_0$, we have
 \begin{equation}\label{alpham}
   \left(\frac{1}{\alpha m}\right)^{\frac{1}{m-1}}=
   \frac{24m}{1+2m}\left[
   1-q_0+\frac{9}{n^2}+\frac{8\pi G\rho_0}{12H_0^2}
   \right]
   H_0^2.
 \end{equation}
 Substituting (\ref{alpham}) into (\ref{ai}), we obtain
  the relationship between parameters $m$ and $n$
 as follows:
 \begin{equation}\label{mvalue}
   m=\frac{5-4q_0+\frac{36}{n^2}}{2(1-2q_0)+\frac{36}{n^2}+
   \frac{8\pi G\rho_0}{H_0^2}},
 \end{equation}
 where $G=6.67\times 10^{-11}kg^{-1}m^3s^{-2},
 \rho_0 = (3.8\pm 0.2)\times 10^{-28}kgm^{-3},
 H_0=(2.3\pm 0.1)\times 10^{-18}s^{-1}$\cite{experiment}
  and therefore
 $\frac{8\pi G\rho_0}{H_0^2}\approx 0.12$.
 One necessary and sufficient condition for the present accelerated
 expansion of the universe is that the deceleration parameter
 $q_0<0$. Specifically, the allowed range for $q_0$ according to
 the present observation is
 $q_0=(-0.57\pm 0.10)$ \cite{experiment}. Here we apply this criteria
 to determine the allowed range for $m$ and $n$, as
 depicted in Fig. \ref{range1}.
 From the figure, it is obvious that there do exist suitable
 parameters $m$ and $n$, that is, suitable formalisms of
 $f(R)=\alpha R^m$ and corresponding relationships between the
 extra dimension and the other ones,
  such that the present cosmic acceleration can be
 explained without the need of dark energy.
 Besides, Fig. 1 also reveals that both positive and
 negative values of $n$ is allowed, indicating that the idea of dynamical
 compactification
 can be extended to that both the expansion and the contraction of
 the extra dimension can be responsible for the accelerated expansion
 of the other dimensions.
 Besides,
   we know from equation (\ref{alpham})
 that
 \begin{equation}\label{alpha}
   \alpha =
    \frac{1}{m}
    \left[\frac{24m}{1+2m}\left(1-q_0+\frac{9}{n^2} +\frac{
    8\pi G\rho_0}{12H_0^2}\right)H_0^2\right]^{-(m-1)}.
 \end{equation}
 \begin{figure}
   \begin{center}
     \includegraphics[height=7cm,width=13cm]{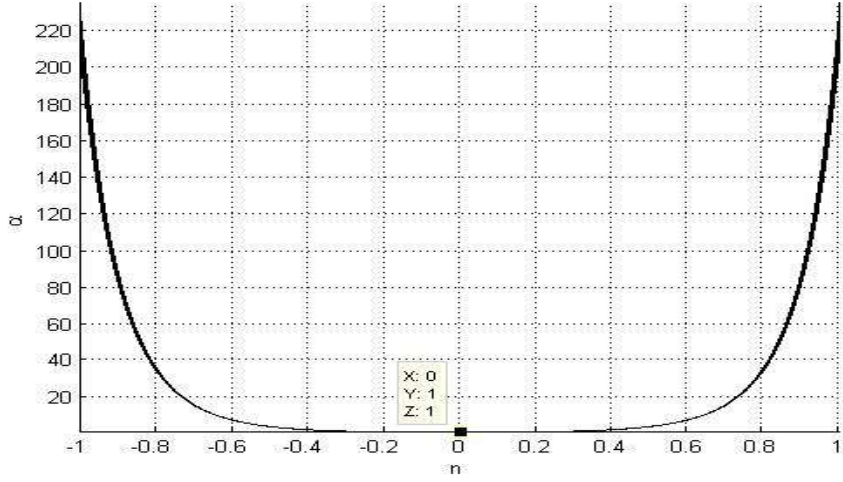}
     \caption{The relation between $n$ and the possible range for
     $\alpha$. The range is spanned by the uncertainty of the
     deceleration parameter $q_0$.}\label{alphan}
   \end{center}
 \end{figure}
 Using Eqs. (\ref{mvalue}) and (\ref{alpha}), we depict the $\alpha - n$ relation
 in Fig. \ref{alphan} with the range spanned by $q_0$, from which we
 find $\alpha$ increases rapidly as $|n|$ increases.
 Further, one could find from Fig. \ref{range1} or equation
 (\ref{mvalue}) that
 the lower bound of $m$ is 1, corresponding to
 $\alpha=1$, and therefore $f(R)=R$, which is exactly the KK
 case. One subtlety concerning this case is that
  Fig. \ref{range1} as well as equation (\ref{mvalue}) shows that
   $n=0$ for $m=1$, and therefore equation(\ref{kk}) apparently
 indicates
 $H_0=\dot a=0$, contradicting
 with the present observations. Thus one might think that
  the KK theory is inconsistent with the observation.
   However, this conclusion is
  misleading, because from the inverse-solving procedure in equation
  (\ref{rphi}), we have already
 confined our discussion to $m\neq 1$ cases.
  On the other hand,
 the asymptotic value of $m$, as $n$ approaches infinity, is
 \begin{equation}\label{masymptotic}
   \lim_{n\rightarrow\infty}m=
   \frac{5-4q_0}{2(1-2q_0)+0.12},
 \end{equation}
  from which we find the upper bound for $m$ to be $1.72$.
 Hence, the possible range for $m$ is
 narrowed down as $1<m<1.72$.

\begin{figure}
\begin{center}
 \parbox{8cm}{
\includegraphics[height=7cm,width=7cm]{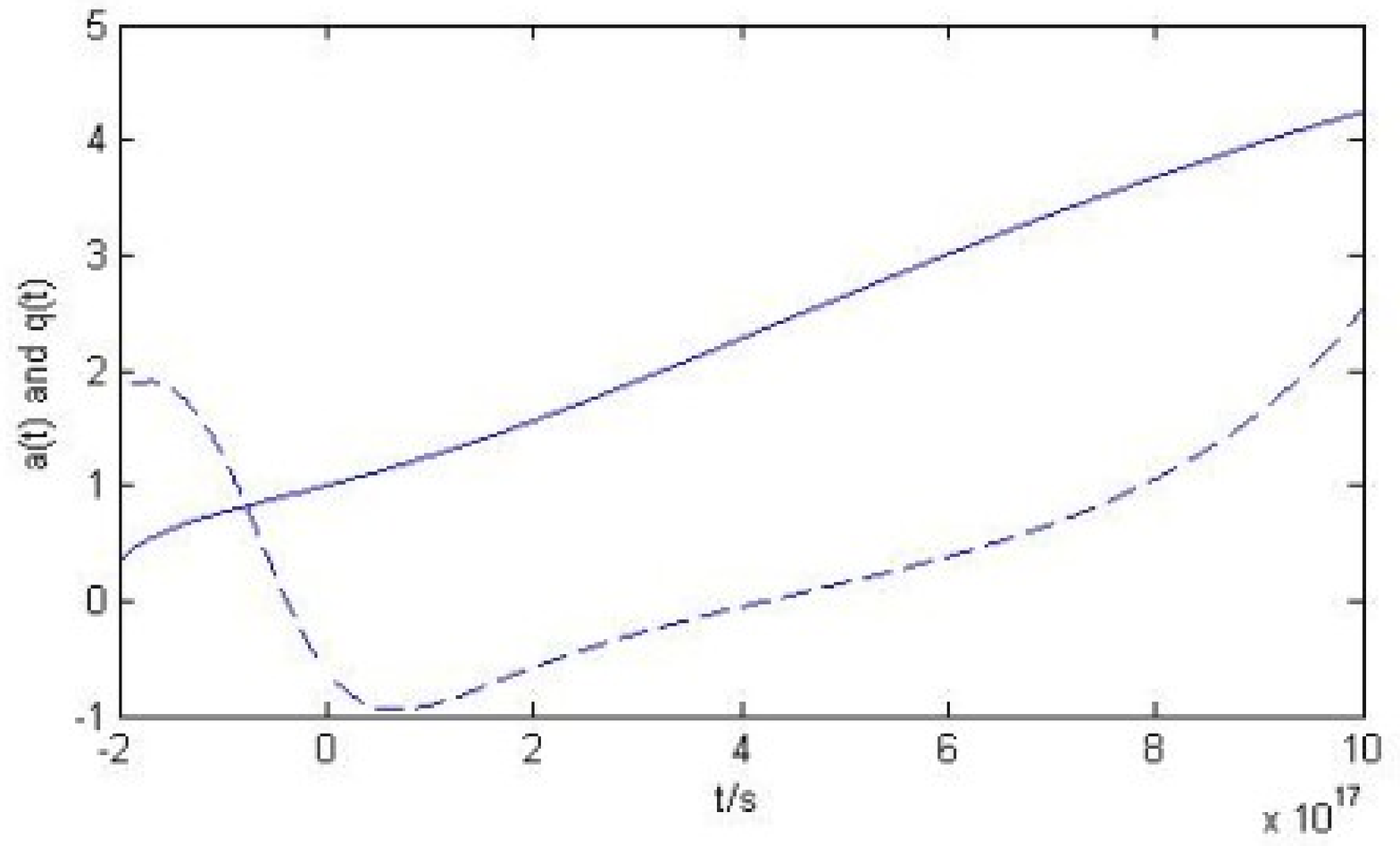}
\parbox{7cm}{
\caption{The evolution of $a(t)$ (solid line) and $q(t)$ (dashed
line) with $m=1.5, n=5.4$. Here t=0 corresponds to today, and the
present value of the deceleration parameter is
$q_0=-0.6$.\label{m15n54q06aq}} } }
\parbox{8cm}{
 \includegraphics[height=7cm,width=7cm]{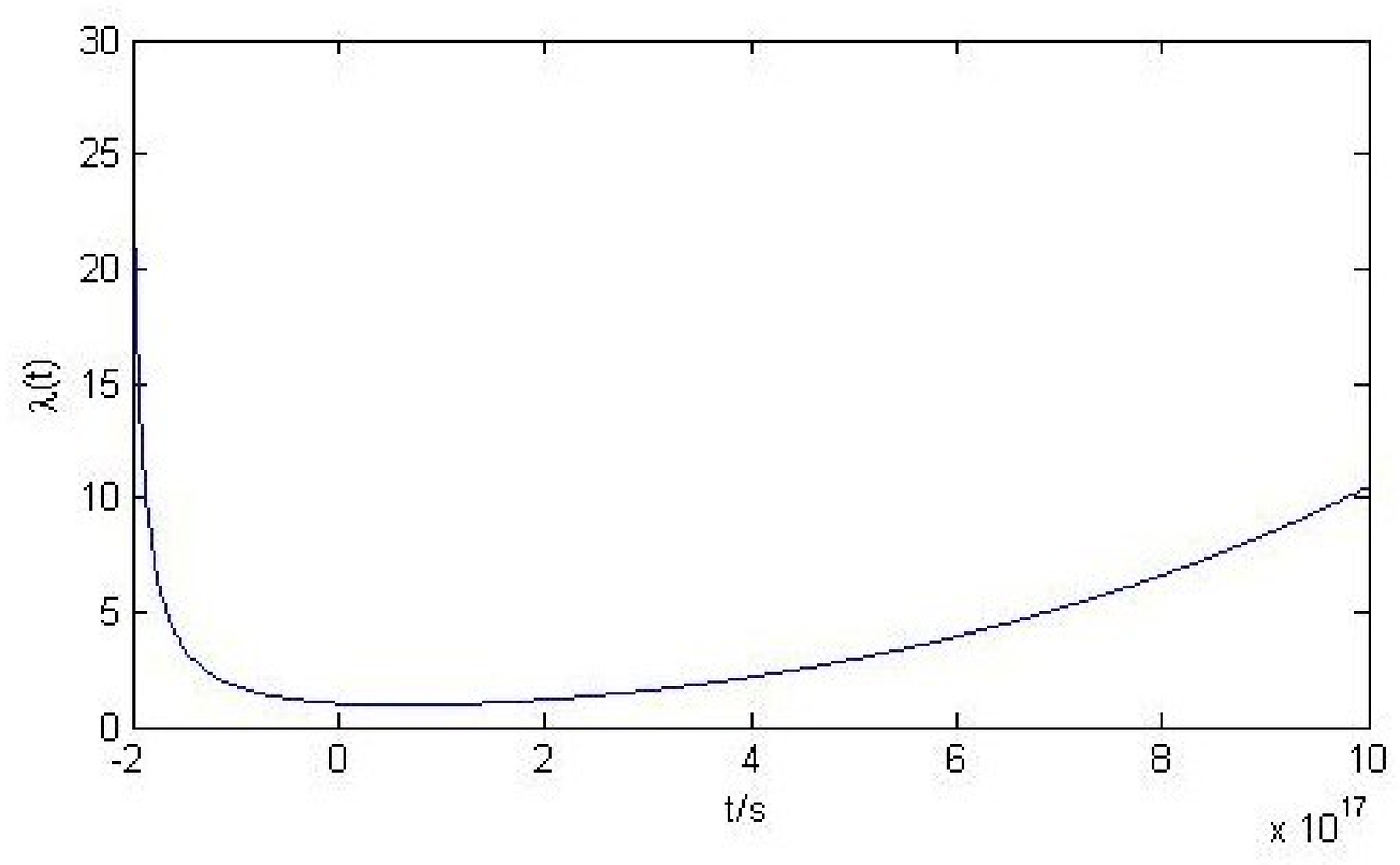}
 \parbox{7cm}{
\caption{The evolution of $\lambda(t)$ with $m=1.5, n=5.4$. Note
that $\lambda(t)$ is decreasing at $t=0$.\label{m15n54q06lambda}} }}
\end{center}
\end{figure}
\begin{figure}
\begin{center}
\parbox{8cm}{
 \includegraphics[height=7cm,width=7cm]{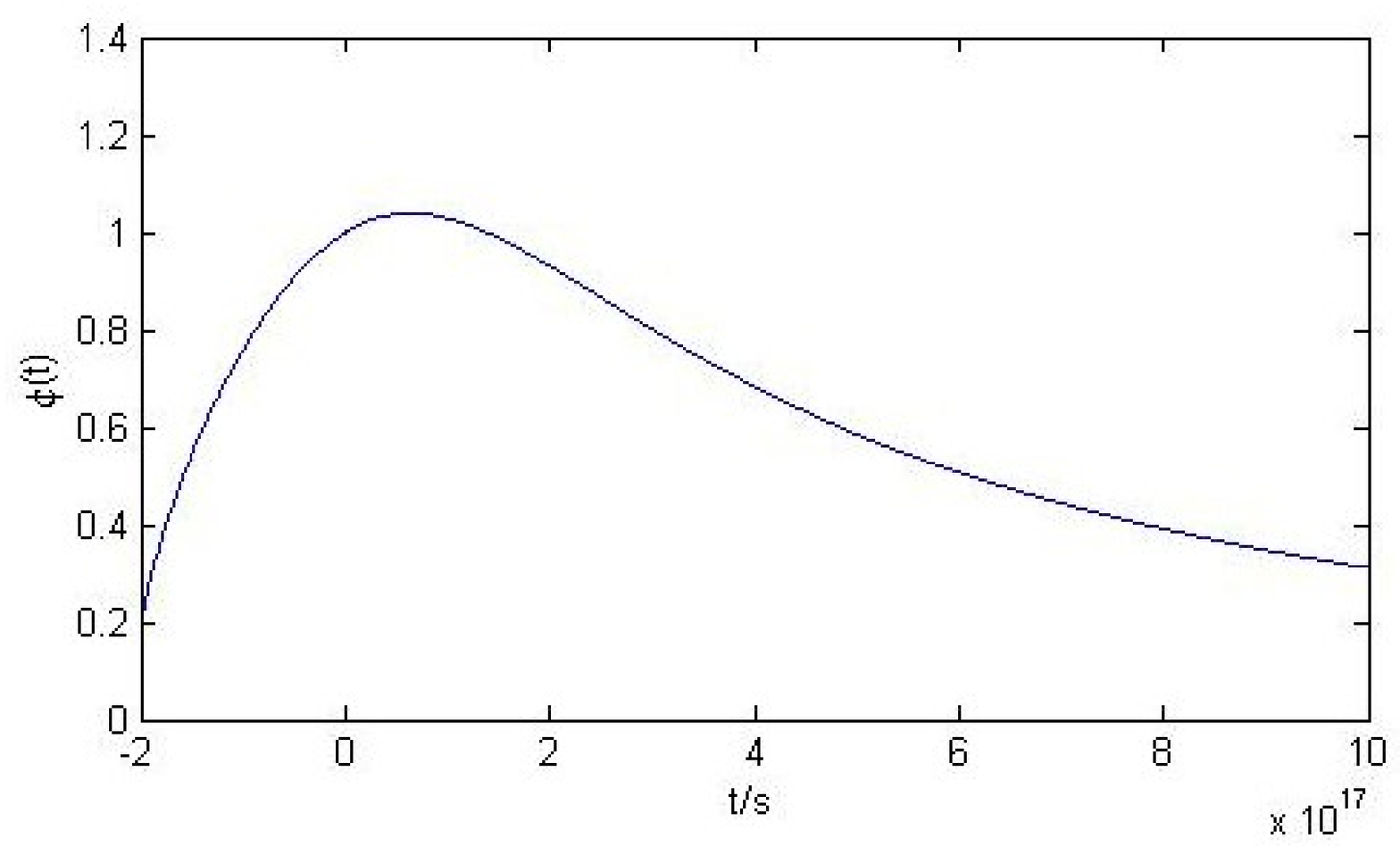}
 \parbox{7cm}{
\caption{The evolution of $\phi(t)$ with $m=1.5, n=5.4$.
\label{m15n54q06phi}} }}
\parbox{8cm}{
  \includegraphics[height=7cm,width=7cm]{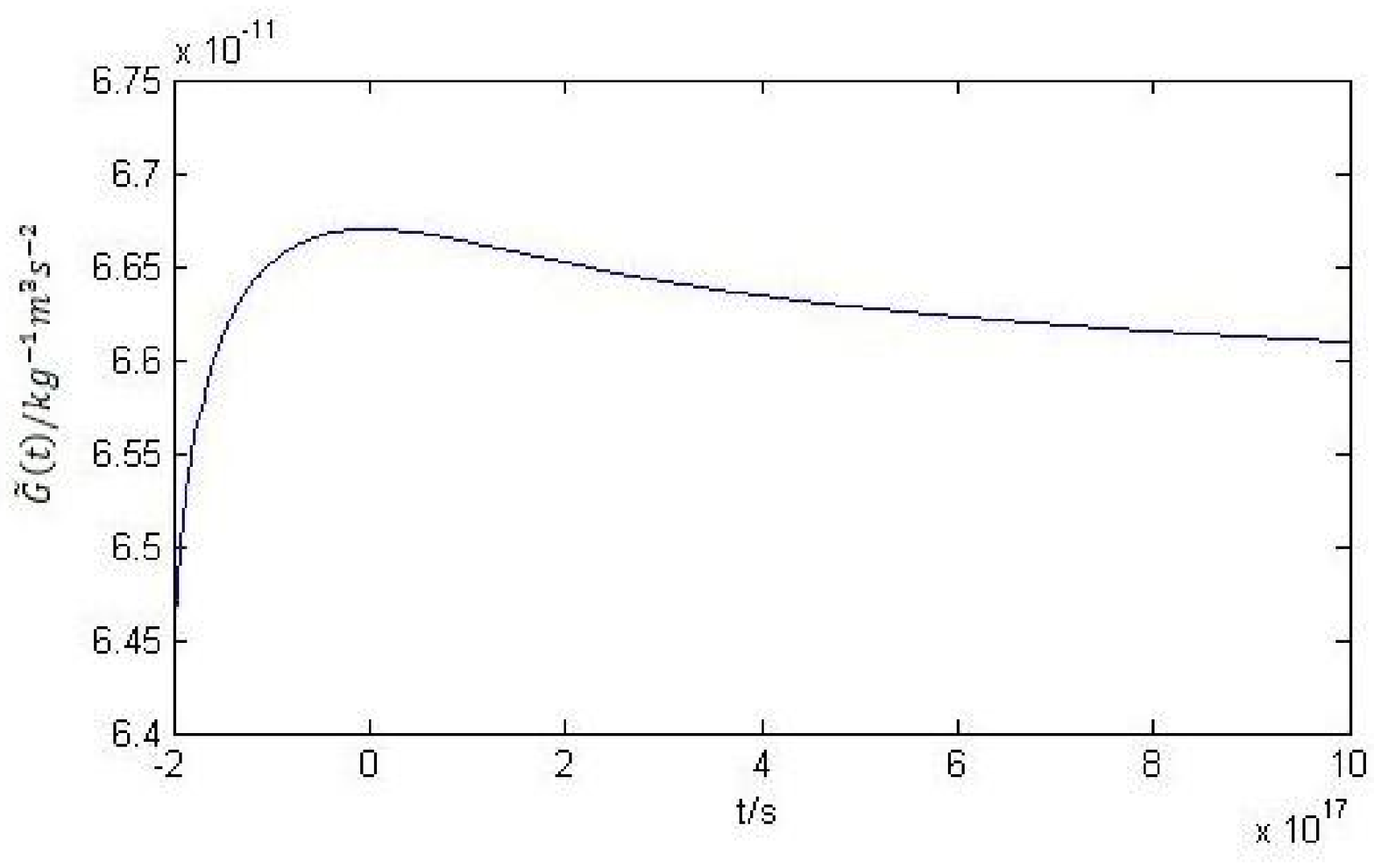}
  \parbox{7cm}{
\caption{The evolution of $\tilde{G}(t)$  with $m=1.5,
n=5.4$.\label{m15n54q06g}}  }}
\end{center}
\end{figure}

\begin{figure}
\begin{center}
 \parbox{10cm}{
\includegraphics[height=7cm,width=10cm]{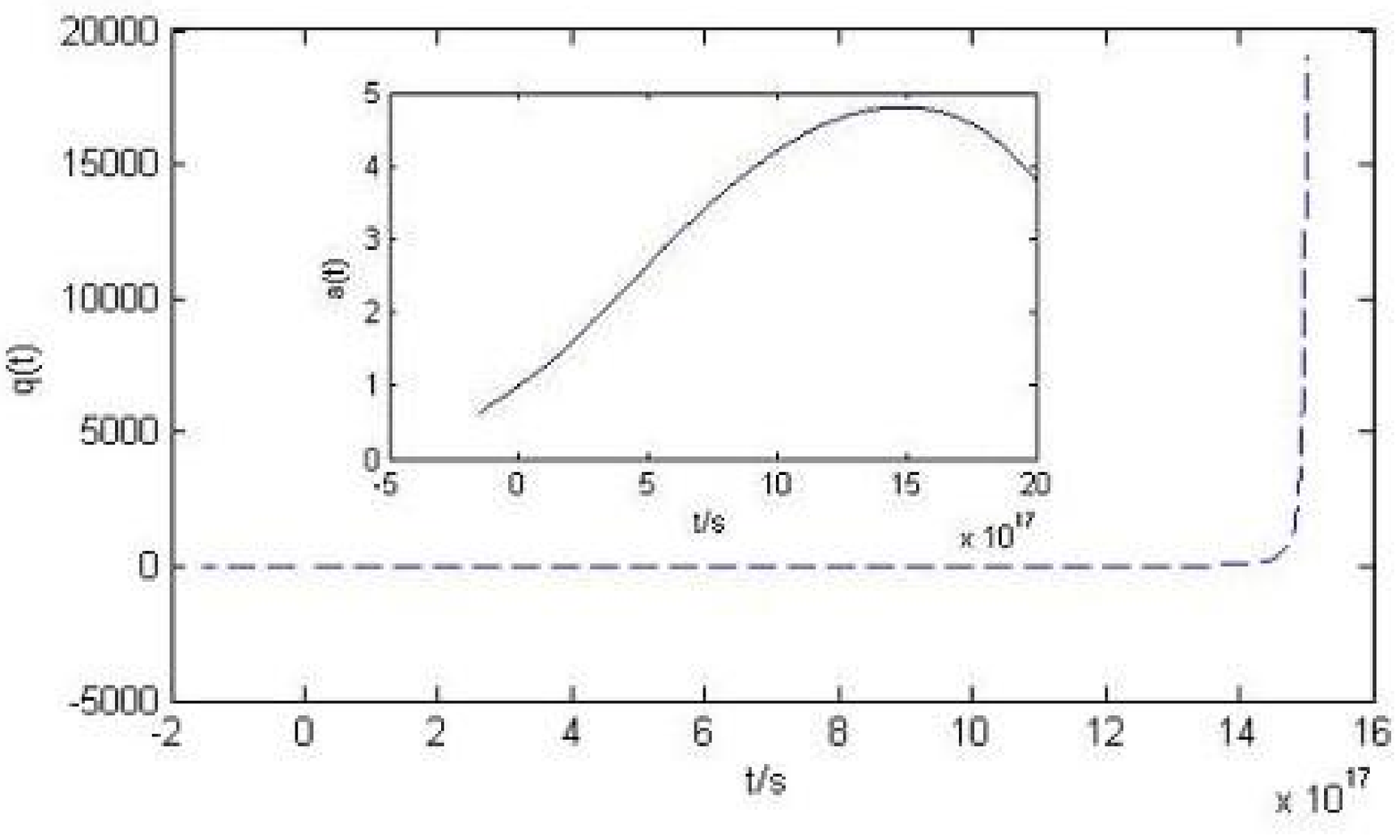}
 \parbox{11cm}{\caption{
 Numerical simulation for a longer period of cosmic
evolution. The deceleration parameter $q(t)$ becomes divergent at
$t=(13.5+47.6)$Gyr(dashed line), as $a(t)$ (solid line in the
subfigure) approaches a constant value.} \label{diverge}
 }
 }
\end{center}
\end{figure}

 With equations (\ref{ai}) -- (\ref{phii}), initial values
 (\ref{ini}) and (\ref{alpham}),
 the numerical simulation of the
 evolution of the scale factor $a(t)$,
 the scalar field $\lambda(t), \phi(t)$ and the deceleration
 parameter $q(t)$ can be conducted.
 As an example, we choose $m=1.5, n=5.4$. Then from equation
 (\ref{mvalue}) we get $q_0=-0.6$.
 The corresponding value of $\left(\frac{1}{\alpha m}\right)
 ^{\frac{1}{m-1}}$ is approximately $17.3H_0^2$. It is
 useful to define
 the effective gravitational ``constant''
  from Eqs. (\ref{4dfield}) -- (\ref{4dfr})
 as
 \begin{equation}
   \tilde{G}=G\lambda^{-1/2}f'(R)^{-1}=G\lambda^{-1/2}\phi^{-1}.
 \end{equation}
 Note that $\tilde{G}$
  is actually an evolving scalar.
 Then the evolution characters
  of $a(t), q(t), \lambda(t), \phi(t)$ and $
 \tilde{G}$ are illustrated in Figs. \ref{m15n54q06aq} --
 \ref{m15n54q06g}  respectively. From these figures,
  it is clear that the present deceleration
  parameter $q_0=-0.6$ is in consistent with observation
  and the above analysis. Moreover, we note the following:
 \begin{enumerate}
   \item
   $q(t)$ rolls from a positive value to a negative one
   smoothly in the recent past. Specifically, if the age of the
   universe is $13.5$Gyr \cite{experiment}, the universe
   turns from deceleration to acceleration at $t=(13.5-1.2)$Gyr.
   Thus, the ``coincidence'' problem
   in dark energy can be addressed in this case.
   \item
   After $q(t)$ reaches the minimal value, it rolls back
   again and becomes positive at $t=(13.5+13.4)$Gyr. Thus,
    the universe will become
   decelerating in the future rather than be endlessly accelerating.
   \item
   Numerical simulation for a long period of cosmic
   evolution shows that the value of $q(t)$ becomes divergent at
   $t=(13.5+47.6)Gyr$, as depicted in Fig.
   \ref{diverge}. Since $a(t)>0$ and its evolution
   is generally slow, the only reason for the divergence of
   $q(t)$ is $\dot a\rightarrow 0$. In other words,
   the universe would come to a static point
    in the finite future.
   Actually in this period, numerical simulation shows that
    all quantities, such as $\lambda(t)$ and
   $\phi(t)$, still remain well-behaved. After this static point,
   $a(t)$ starts to decrease for a certain period before numerical
   simulation fails. Thus, around the static point of $a(t)$,
   numerical simulation is still effective.
   \item
   $\lambda(t)$ decreases from a large value in the past and
   increases slowly at present, while $\phi(t)$ increases to a
   maximum value for the present period
   and decreases slowly in the future.
   After the static point of $a(t)$, nonetheless, the decrease of
   $\phi(t)$ becomes faster and therefore it would approach $0$
   in the finite future when the numerical simulation fails.
   Actually, since $\phi$ appears in the denominators of Eqs.
   (\ref{ai}) -- (\ref{phii}), the analysis from this set of
   equations is no longer applicable for the evolution character
   beyond the $\phi=0$ point. Instead, one should take advantage of
   original equation (\ref{5dfield}) (or Eqs. (\ref{4dfield})
   -- (\ref{4dfr}) by multiplying $f'(R)$ on both sides). We
   leave the details about the static point and the
   evolution beyond to future investigations.
   \item The evolution of $\tilde{G}$ is generally slow in the
   future. The present value of $\tilde{G}$ is maximal with
   an almost zero evolving speed.
   Thus, it is also in consistent with the observation data that
   the present gravitational constant has a negligible evolving
   speed.
    \end{enumerate}

On the other hand, as indicated by Fig. \ref{range1}, the expansion
of
 the extra dimension would also bring about the acceleration of other
  dimensions. If we take $m=1.5, n=-5.4$ instead of $n=+5.4$,
 the numerical simulation will
 give similar evolution features of
 $a(t), q(t)$ and $\tilde{G}(t)$ with
 the present value of deceleration parameter $q_0=-0.6$, which are
 consistent with observations.
 In this case,
 the universe turned from deceleration to acceleration
 at $t=(13.5-6.1)$Gyr, which was $4.9$Gyr earlier than that in the
 $n=5.4$ case.
 Correspondingly,
  the future transition point of the universe from acceleration to
 deceleration is also moved ahead to $t=(13.5+8.0)$Gyr, rendering
 that the
 total acceleration period of the universe changes only moderately
 from $14.6$Gyr to $14.1$Gyr.
 Thus, the $n<0$ case is very similar to the $n>0$ case,
 except that the cosmic
 transition from deceleration to acceleration and
 the inverse occur both ahead of time.
 The reason for such temporal translation can be viewed in the
 following way. As shown before,
 when $1<m<1.72$,
 the future asymptotic evolution of $\lambda$ should be
 increasing. Therefore, if the initial (present) evolution rate for
  $\lambda$ is set to be increasing at present rather than be
  decreasing,
 there should be a temporal translation with respect to the evolution
 of $\lambda$ and therefore to other quantities that can be
 affected.

\section{Discussions}\label{jielun}

\begin{figure}
\begin{center}
\parbox{8cm}{
\includegraphics[height=6cm,width=7cm]{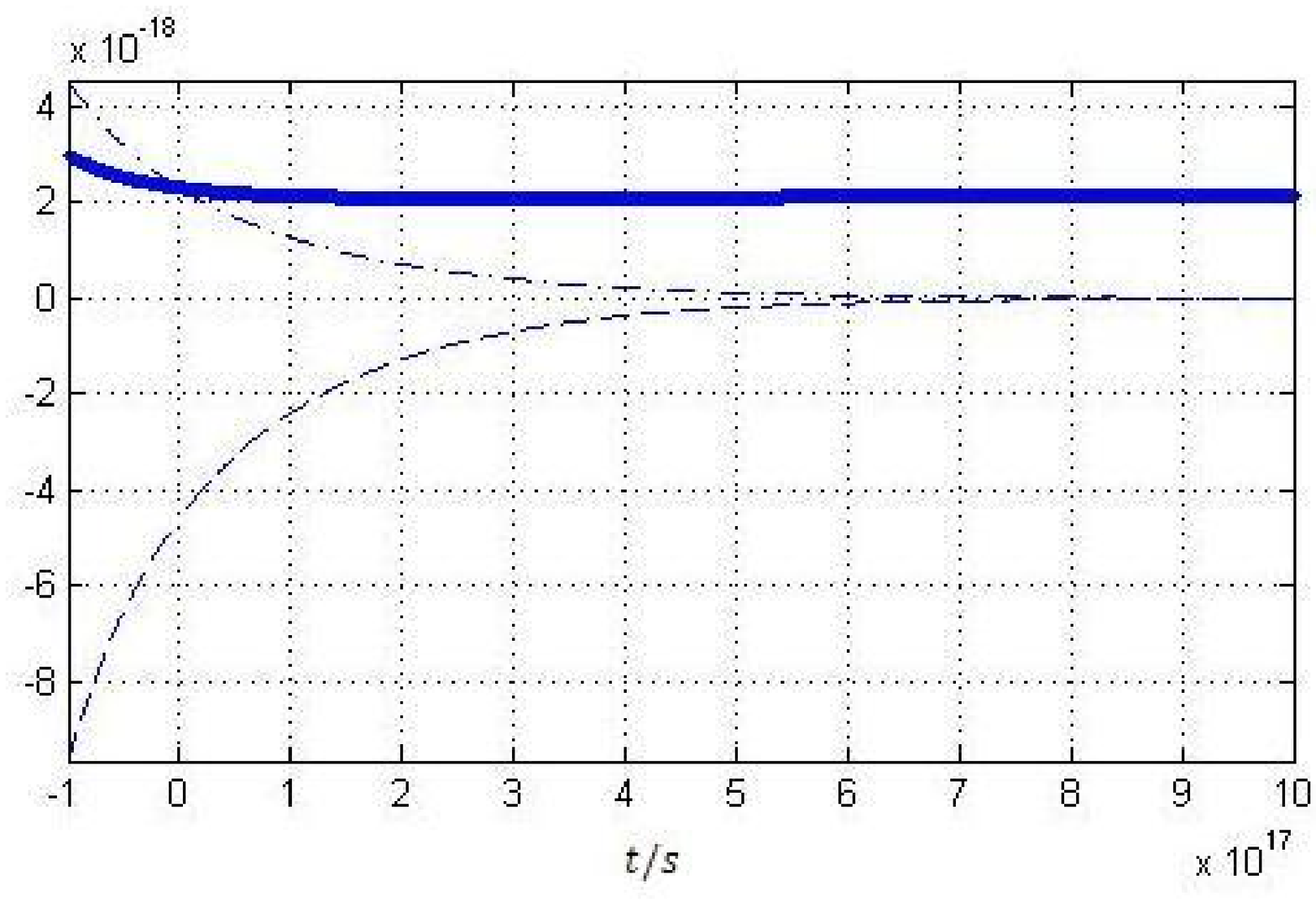}
\parbox{7cm}{
\caption{The evolution rate of $\daa$ (solid line), $\dll$ (dashed
line) and $\dpp$ (dash-dotted line) in the 5-D BD theory, with
parameters $\omega=1.2, n=3$ in \cite{5dbd}. } \label{ratebd} }}
\parbox{8cm}{
\includegraphics[height=6cm,width=7cm]{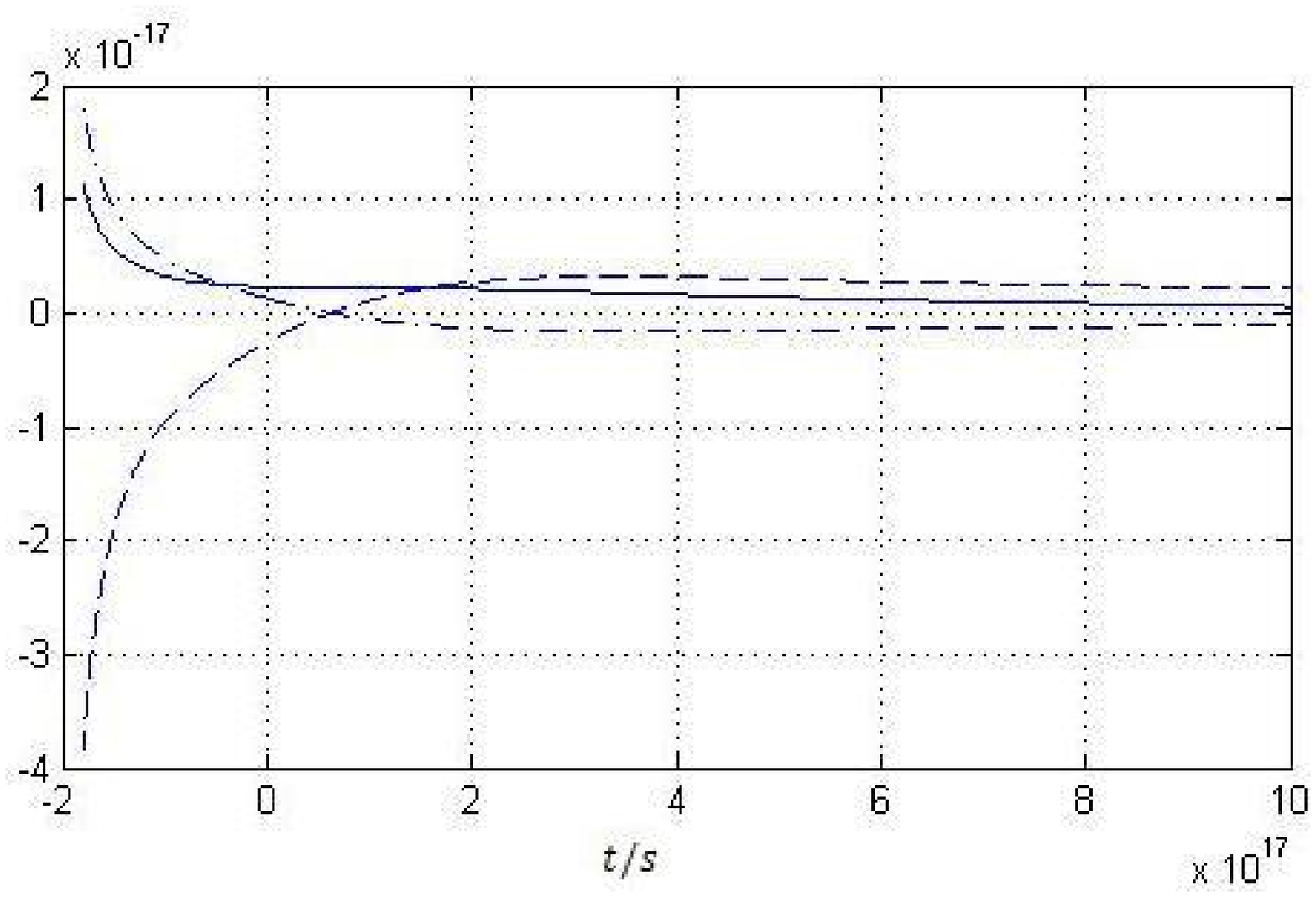}
\parbox{7cm}{
 \caption{
The evolution of $\daa$ (solid line), $\dll$ (dashed line) and
$\dpp$ (dash-dotted line) in 5-D $f(R)$ gravity with parameters
$m=1.5, n=5.4$.  \label{ratefr}} }}
 \end{center}
 \end{figure}

 To clarify the argument for the
  equivalence
  of the $f(R)$ theory of gravity to
  certain special cases of BD theory up to
  a potential term \cite{bdfr}, we now compare the cosmic evolution of 5-D $f(R)$ models developed
 in previous sections to the
 5-D BD theory \cite{5dbd}.
 From equation (\ref{4dfield}), it is noticeable
 that the potential term is $\frac{f(R)-R f'(R)}{f'(R)}$, and similar
 terms also exist in equations (\ref{4dlambda}) and (\ref{4dfr}).
  These terms correspond to the $\pot$ terms
  in (\ref{ai}) -- (\ref{phii}), and would be shown to
  play significant roles in
 predicting the future evolution of the universe.
 There are three independent functions in the
 5-D BD theory \cite{5dbd}, namely
 $a(t)$, $\lambda(t)$ and $\phi(t)$, where $\phi(t)$ represents the
 core idea of BD theory on an evolving gravitation ``constant'' due to
 the interaction between
  the local gravitational fields with the faraway
 matter in the universe.
  We note that $q(t)$ in this case
  (see Fig. 2 in \cite{5dbd}) evolves to a constant in the
  future. This means that the universe would accelerate
 permanently in the future until its energy density decreases to be
 extremely low so that certain quantum effects may become significant,
 such as the situation in \cite{loop}.
 It can also be read from Figs. 3 -- 5 in \cite{5dbd}
  that the asymptotic values of
 $\dll$ and $\dpp$ are both zero, as further illustrated in
 Fig. \ref{ratebd}.

  In contrast, in our 5-D $f(R)$ theories,
  Fig. \ref{m15n54q06phi} has demonstrated that, while
 $\phi$ evolves to
 a large value presently and remains positive in the future,
  the $\pot$ terms render the evolution of
 $\lambda(t), \phi(t)$ to be more complicated.
  Specifically, the corresponding evolution characteristics of
  $\daa, \dll, \dpp$ are illustrated in Fig. \ref{ratefr}
  with specific parameters $m=1.5, n=5.4$.
  As $\tilde{G}=G\lambda^{1/2}\phi^{-1}$ is an observable, it
 is interesting to show how $\pot$ terms affect its evolution.
  Since $m>1$, $q_0<0$, from equation
 (\ref{alpham}) we have
 $\left(\frac{1}{\alpha m}\right)^{\frac{1}{m-1}} > 0$.
 Moreover, we do not consider
  any solution with a negative $\phi$
  because of its anti-gravity character.
 Thus, the coefficient of the $\pot$ term would determine the
 evolutional inclination of $\lambda$ and $\phi$.
 Specifically, because $m=0.5$ and $m=2.5$ correspond to
 $\left(\frac{1}{4m}-\frac{1}{2}\right)=0$
 and $\left(\frac{5}{8m}-\frac{1}{4}\right)=0$ respectively,
 the qualitative features of the future evolution
 of $\lambda$ and $\phi$ are as follows:
 \begin{itemize}
   \item If $m<0.5$, $\lambda \downarrow,\quad \phi \downarrow$
   \item If $m>2.5$, $\lambda \uparrow,\quad \phi \uparrow$
   \item If $0.5<m<2.5$, $\lambda \uparrow,\quad
   \phi \downarrow$
 \end{itemize}
 In view of the constraint illustrated by Fig. \ref{range1},
 only the last situation is allowed, where the evolution of
  $\lambda$ and $\phi$ counterbalances each other and
  therefore the evolution
  of $\tilde{G}$ is generally slow.

In summary, we present the Killing reduction of 5-D $f(R)$ theories
of gravity
  to the 4-D sensible world.
  Then we study its cosmological
 implication by assuming the spatial homogeneity and isotropy, namely,
 the Friedman-Robertson-Walker metric.
 With the illustration of a specific example
 it is found that the theory is consistent with
 the present observations in a variety of aspects,
  including the recent cross from cosmic
 deceleration to acceleration, its prescription
 of present cosmic speed-up, the negligible evolving speed of the
 effective gravitational constant, and etc..
 In contrast to the KK cosmology,
  it is worth mentioning that in 5-D $f(R)$ theories of
 gravity, both expansion and contraction of the extra dimension
 could result in the present
  accelerated expansion of other spatial dimensions.
 Even for a simple and generic class of $f(R)=\alpha R^m$ models, the
 5-D $f(R)$ theories of gravity do not need unreasonable or
 ill-initiated fine-tuning of certain parameters.
  Hence it is reasonable to infer that the present accelerated
  expansion of spatial dimensions could be certain basic character
 of the 5-D spacetimes.

 Finally, it is pointed out in Ref.
  \cite{loop} that in the
 large scale limit, quantum gravity effect would come to be vital
 and bring about
  higher-order quantum corrections to the Friedman equation
 to prescribe a static point in the finite future,
 followed by the recollapse of the universe.
 Note that the static point in the finite future is also a
 significant character of the 5-D $f(R)$ gravity.
 Thus, it is
 of interest to explore whether there are
 some relation between the 5-D theories of $f(R)=\alpha R^m$ models
 and the quantum gravity theory applied in \cite{loop}.
 In addition, the future recollapse also
  differs from
 the prediction of endless acceleration scenario in the
 5-D BD cosmology \cite{5dbd}. Hence, 5-D $f(R)$
 gravity is not simply equivalent to 5-D BD theory, since
 the effect of potential terms are shown to be vital enough to
 prescribe
 different cosmological scenarios in the distant future.

 \section*{Acknowledgements}
 The authors would like to thank Li-e Qiang for valuable
  discussions. This
 work is a part of projects 10675019 and 10975017 supported by NSFC.
 Biao Huang would also like to acknowledge support from the Beijing
 Undergraduate Research and Entrepreneurship Foundation.

 \end{document}